\documentclass[sigconf,natbib=false,nonacm]{acmart}
\AtBeginDocument{%
  }

\copyrightyear{2026}
\acmYear{2026}
\setcopyright{cc}
\setcctype{by}
\acmConference[SAC '26]{The 41st ACM/SIGAPP Symposium on Applied Computing}{March 23--27, 2026}{Thessaloniki, Greece}
\acmBooktitle{The 41st ACM/SIGAPP Symposium on Applied Computing (SAC '26), March 23--27, 2026, Thessaloniki, Greece}
\acmPrice{}
\acmDOI{10.1145/3748522.3779992}
\acmISBN{979-8-4007-2294-3/2026/03}

\usepackage{amsmath,amsfonts}
\usepackage{algorithmic}
\usepackage{graphicx}
\usepackage{textcomp}
\usepackage{xcolor}
\usepackage{xspace}
\usepackage{hyperref}
\usepackage{listings}
\usepackage{enumitem}
\usepackage{multirow}
\usepackage{booktabs}
\usepackage{array}
\usepackage{subcaption}
\usepackage{pifont}
\usepackage{fancyvrb}
\usepackage{optidef}
\usepackage{makecell}

\usepackage{listings}

\usepackage{multicol}
\usepackage{adjustbox}
\usepackage{colortbl}
\usepackage{tabstackengine}

\newcommand{\codeinline}[1]{\lstinline[language=python,basicstyle=\ttfamily,breaklines=true,breakatwhitespace=true,showstringspaces=false]{#1}}

\usepackage[acronym]{glossaries}

\hypersetup{%
  colorlinks=true,
  linkcolor=black,
  citecolor=black,
  urlcolor=black
}

\newacronym{ML}{ML}{Machine Learning}
\newacronym{DL}{DL}{Deep Learning}

\graphicspath{{figures/}}

\newcolumntype{C}[1]{>{\centering\let\newline\\\arraybackslash\hspace{0pt}}m{#1}}

\RequirePackage[
  datamodel=acmdatamodel,
  style=acmnumeric,
  ]{biblatex}

\begin{document}

\title{SourceBroken: A large-scale analysis on the (un)reliability of SourceRank in the PyPI ecosystem}
\renewcommand{\shorttitle}{SourceBroken}

\author{Biagio Montaruli}
\email{biagio.montaruli@eurecom.fr}
\affiliation{%
  \institution{EURECOM \& SAP}
  \city{Nice}
  \country{France}
}

\author{Serena Elisa Ponta}
\email{serena.ponta@sap.com}
\affiliation{%
  \institution{SAP}
  \city{Mougins}
  \country{France}
}

\author{Luca Compagna}
\email{lcompagna@endor.ai}
\affiliation{%
  \institution{Endor Labs}
  \city{Palo Alto}
  \country{USA}
}

\author{Davide Balzarotti}
\email{davide.balzarotti@eurecom.fr}
\affiliation{%
  \institution{EURECOM}
  \city{Biot}
  \country{France}
}

\renewcommand{\shortauthors}{B. Montaruli et al.}

\newcommand{\sota}{state-of-the-art\xspace}
\newcommand{\diff}[2]{\frac{\partial #1}{\partial #2}}
\newcommand{\vct}[1]{\ensuremath{\boldsymbol{#1}}}
\newcommand{\mat}[1]{\ensuremath{\mathbf{#1}}}
\newcommand{\set}[1]{\ensuremath{\mathcal{#1}}}
\newcommand{\con}[1]{\ensuremath{\mathsf{#1}}}
\newcommand{\tsum}{\ensuremath{\textstyle \sum}}
\newcommand{\T}{\ensuremath{\top}}
\newcommand{\ind}[1]{\ensuremath{\mathbbm 1_{#1}}}
\newcommand{\argmax}{\operatornamewithlimits{\arg\,\max}}
\newcommand{\erf}{\text{erf}}
\newcommand{\sign}{\text{sign}}
\newcommand{\argmin}{\operatornamewithlimits{\arg\,\min}}
\newcommand{\bmat}[1]{\begin{bmatrix}#1\end{bmatrix}}
\newcommand{\myparagraph}[1]{\noindent \textbf{#1}}
\newcommand{\myparagraphlb}[1]{\noindent \newline \textbf{#1}}
\newcommand{\mysubparagraph}[1]{\noindent \underline{\textit{#1}}}
\newcommand{\ie}{i.e.\xspace}
\newcommand{\eg}{e.g.\xspace}
\newcommand{\etc}{etc.\xspace}
\newcommand{\aka}{a.k.a.\xspace}
\newcommand{\wrt}{w.r.t.\xspace}
\newcommand{\etal}{\emph{et al.}\xspace}

\newcommand{\advms}{AdvModSec\xspace}
\newcommand{\mlms}{MLModSec\xspace}
\newcommand{\modsecurity}{ModSecurity\xspace}
\newcommand{\crs}{CRS\xspace}
\newcommand{\mscrs}{MS-CRS\xspace}
\newcommand{\wafamole}{WAF-A-MoLE\xspace}
\newcommand{\sqligot}{SQLiGoT\xspace}

\newcommand{\red}[1]{\textcolor{red}{#1}}

\newcommand{\mynote}[2]{\xspace\fbox{\bfseries\sffamily\footnotesize{#1}}
{\scriptsize$\blacktriangleright$\textsf{\small\emph{#2}}$\blacktriangleleft$}}
\newcommand{\mybluenote}[2]{\textcolor{blue}{\mynote{#1}{#2}}}
\newcommand{\bm}[1]{\mybluenote{Biagio}{#1}}
\newcommand{\se}[1]{\mybluenote{Serena}{#1}}
\newcommand{\db}[1]{\mybluenote{Davide}{#1}}
\newcommand{\lc}[1]{\mybluenote{LucaC}{#1}}

\newcommand{\takeaways}[1]{
\noindent \fcolorbox{gray}{white}{\begin{minipage}[c]{0.96\columnwidth}
\textbf{Takeaways}: #1
\end{minipage}}
}
\newcommand{\takeawaysNoTitle}[1]{
\noindent \fcolorbox{gray}{white}{\begin{minipage}[c]{0.96\columnwidth}
#1
\end{minipage}}
}

\begin{abstract}
SourceRank is a scoring system made of 18 metrics that assess the popularity and quality of open-source packages.
Despite being used in several recent studies, none has thoroughly analyzed its reliability against evasion attacks aimed at inflating the score of malicious packages, thereby masquerading them as trustworthy.
To fill this gap, we first propose a threat model that identifies potential evasion approaches for each metric, including the \emph{URL confusion} technique, which can affect 5 out of the 18 metrics by leveraging a URL pointing to a legitimate repository potentially unrelated to the malicious package.

Furthermore, we study the reliability of SourceRank in the PyPI ecosystem by analyzing the SourceRank distributions of benign and malicious packages in the state-of-the-art MalwareBench dataset, as well as in a real-world dataset of 122,398 packages.
Our analysis reveals that, while historical data suggests a clear distinction between benign and malicious packages, the real-world distributions overlap significantly, mainly due to SourceRank's failure to timely reflect package removals.
As a result, SourceRank cannot be reliably used to discriminate between benign and malicious packages in real-world scenarios, nor to select benign packages among those available on PyPI.

Finally, our analysis reveals that \emph{URL confusion} represents an emerging attack vector, with its prevalence increasing from 4.2\% in MalwareBench to 7.0\% in our real-world dataset.
Moreover, this technique is often used alongside other evasion techniques and can significantly inflate the SourceRank metrics of malicious packages.
\end{abstract}



\maketitle

\section{Introduction}
Supply chain attacks are a growing threat to the software industry.
According to a report released by Sonatype in 2024,
the number of malicious packages discovered in the wild had a yearly increase of 156\%, with 512,847 new malicious packages identified in 2024~\cite{sonatype2024}.
Recently, PyPI, one of the main ecosystems for Python packages, has faced a growing number of attacks that even led to a temporary halt in the creation of new projects and the registration of new users~\cite{Checkmarx2024BlogPost,Goodin2024ArsTechnicaArticle}.

To ensure the reliability of the packages available in the ecosystem and automate decisions on the
trustworthiness of open-source projects,
several scoring systems have been proposed, such as SourceRank~\cite{Sun2023SoftwareProvenance,Saini2020SoftwareMetrics,sourcerank_blog} and the OpenSSF Scorecard~\cite{scorecard_paper,OpenSSFScorecard}.
In this work, we focus on SourceRank, a well-known scoring system developed by Libraries.io for ranking open-source software packages,
which combines 18 metrics, such as number of dependent packages, frequency of releases, number of contributors, and presence of informative artifacts like README file, license, or repository.

The SourceRank score has been adopted in several recent studies to validate the reliability of open-source packages recommended by Large Language Models (LLMs)-based systems such as ChatGPT and Copilot~\cite{latendresse2025robustllmgeneratedlibraryimports,latendresse2025softwarelibrarian,latendresse2024chatgptgoodsoftwarelibrarian},
to rank and select PyPI packages~\cite{Ladisa2023ACSAC,Sun2023SoftwareProvenance,Jia2024EmpiricalStudy},
to serve as a feature for building a prediction model~\cite{Jafari2023DependencyUpdateStrategies}, as well as to infer dependencies of third-party packages~\cite{Ye2022ICSE}.
However, none of the previous studies have thoroughly analyzed the security and reliability of SourceRank against potential evasion attacks
that could be exploited by malicious actors to manipulate the metrics, thereby increasing the SourceRank score of a malicious package and making it appear trustworthy.

In this study, we aim to fill this gap by proposing a threat model of the SourceRank score and analyzing its behavior for benign and malicious packages in the PyPI ecosystem.
For each metric contributing to the SourceRank score, we identify potential evasion approaches aimed at manipulating the metric to
increase the score of a malicious project, thereby making it appear trustworthy.
In particular, we found a novel evasion approach, which we name \emph{URL confusion}, that consists in leveraging the URL of a legitimate package's repository even though the malicious package itself is not related to the referenced URL.
Given that such a URL influences 5 out of the 18 metrics used to compute the SourceRank score, this evasion technique can significantly increase the SourceRank of a malicious package by exploiting the URL of a legitimate one.

We then perform a study on the PyPI ecosystem by analyzing the SourceRank distributions of benign and malicious packages from the state-of-the-art MalwareBench dataset~\cite{malwarebench},
as well as from a large dataset of 122,398 newly uploaded packages collected daily from PyPI over 80 days.
Our analysis shows that while on historical data (\ie, MalwareBench) there is a clear separation between the distributions of benign and malicious packages, in a real-world scenario they significantly overlap due to the fact that SourceRank does not promptly reflect the removal of malicious packages from PyPI. In fact, only 6 out of the 210 malicious packages in the real-world dataset were marked as removed after five months.

Our results highlight that the SourceRank score is not effective in discriminating benign from malicious packages in a
real-world scenario as the best achievable F1-score based on real-world data is only 14.87\%.
Moreover, since SourceRank can only discriminate malicious packages once they are removed from PyPI and once such information is propagated to Libraries.io (information which is often delayed), it is unreliable for selecting benign packages from those available on PyPI.

We finally analyze the presence of the \emph{URL confusion} evasion approach in the datasets, and we show that, even though it is not widespread (7.0\% of the packages in the real-world dataset),
it is an emerging attack technique (\ie, was 4.2\% of the packages in the MalwareBench dataset from 2024) that can significantly increase the SourceRank of a malicious package.

The remainder of the paper is organized as follows.
Section~\ref{sec:sr_threat_modeling} first introduces the SourceRank score and then presents its threat model, including the evasion techniques we identified for each metric that contributes to SourceRank.
Section~\ref{sec:experiments} describes the datasets we used for our experiments, the research questions we aim to answer and the results of our experiments.
Finally, Section~\ref{sec:conclusions} concludes the paper and summarizes the main findings of our study.
\section{SourceRank Threat Modeling}
\label{sec:sr_threat_modeling}
This section first provides an overview of SourceRank, and then presents its threat model
describing how an attacker can manipulate it to make a malicious package appear as trustworthy.

\begin{table*}[!t]
    \centering
    \caption{Summary of the metrics that make up the SourceRank score.
    For each metric, we provide a brief description and how an attacker can manipulate it to increase the SourceRank of a malicious package.
    The metrics highlighted in green are not shown on the Libraries.io website but are provided by the Libraries.io API and used to compute the SourceRank score.
    }
    \begin{adjustbox}{max width=\textwidth}
    \begin{tabular}{c c c}
    \toprule
    \textbf{Metric} & \textbf{Description}  & \textbf{Attack} \\
    \toprule
    \makecell{Basic info \\ present} & \makecell{Whether the package has basic information: \\ description, homepage or repository URL, and keywords.} & 
        \makecell{Add basic info: an appealing description, commonly-used keywords, \\ and a URL pointing to a repository related to a legitimate package \\ (\emph{URL confusion}) or to a new one created by the attacker.} \\
    \midrule
    \multirow{3}{*}{\makecell{Source repository \\ present}} & \multirow{3}{*}{\makecell{Whether the package provides a URL \\ referring to a repository.}} & 
        \makecell{Use \emph{URL confusion}: add a URL pointing to a repository \\ related to a legitimate package.} \\
        \cmidrule{3-3}
        & & \makecell{Create a new repository for the malicious package \\ and provide its URL.} \\
    \midrule
    \makecell{Readme present} & \makecell{Whether the related repository has a README file \\ with information about the package.} & \makecell{Use \emph{URL confusion} or add a README file in repository \\ related to the malicious package.} \\
    \midrule
    \makecell{Has multiple \\ versions} & \makecell{Whether the package has more than one release version.} & \makecell{Create multiple releases of the package, \\ even with legitimate code at first.} \\
    \midrule
    \makecell{Follows SemVer} & \makecell{Whether the package follows Semantic Versioning (SemVer) \\ for its releases.} & \makecell{Follow the SemVer specification for the releases.} \\
    \midrule
    \makecell{Recent release} & \makecell{Whether the package has a release in the last six months.} & \makecell{Create a new release every 6 months, \\ even with legitimate code at first.} \\
    \midrule
    \makecell{Not brand new} & \makecell{Whether the package has existed for at least six months.} & \makecell{Create a legitimate version of the package and wait \\ for six months before updating it with malicious code.} \\
    \midrule
    \makecell{1.0.0 or greater} & \makecell{Whether the package has a release version \\ greater or equal to 1.0.0.} & \makecell{Create a release with a version greater or equal to 1.0.0.} \\
    \midrule
    \makecell{Dependent \\ packages} & \makecell{The number of projects (packages) that depend on this package, \\ scaled logarithmically in base 10 and multiplied by two.} & \makecell{Create many other packages that depend on the malicious one.} \\
    \midrule
    \makecell{Dependent \\ repositories} & \makecell{The number of repositories provided for packages \\ depending on this package, scaled logarithmically in base 10.} & \makecell{Create a repository for each dependent package.} \\
    \midrule
    \multirow{3}{*}{\makecell{Stars}} & \multirow{3}{*}{\makecell{The number of stars the repository provided in the package \\ has received, scaled logarithmically in base 10.}} & 
        \makecell{Use \emph{URL confusion}. The malicious package inherits \\ the stars of the legitimate one.} \\
        \cmidrule{3-3}
        & & \makecell{Create fake accounts to star the repository \\ related to the malicious package.} \\
    \midrule
    \multirow{3}{*}{\makecell{Contributors}} & \multirow{3}{*}{\makecell{The number of contributors to the repository provided in the package, \\ scaled logarithmically in base 10 and divided by two.}} & 
        \makecell{Use \emph{URL confusion}. The malicious package inherits \\ the contributors of the legitimate one.} \\
        \cmidrule{3-3}
        & & \makecell{Create fake accounts that contribute to the repository \\ provided in the malicious package.} \\
    \midrule
    \makecell{Libraries.io \\ Subscribers} & \makecell{The number of subscribers to the package on Libraries.io, \\ scaled logarithmically in base 10 and divided by two.} & \makecell{Create fake accounts that subscribe to the package on Libraries.io.} \\
    \midrule
    \multicolumn{1}{>{\columncolor{green!40}}c}{\tabularCenterstack{c}{All pre-releases}} & \multicolumn{1}{>{\columncolor{green!40}}c}{\tabularCenterstack{c}{Whether the package has all pre-release versions, \\ such as alpha or beta releases.}} & \multicolumn{1}{>{\columncolor{green!40}}c}{\tabularCenterstack{c}{Avoid creating pre-release versions of the malicious package.}} \\
    \midrule
    \multicolumn{1}{>{\columncolor{green!40}}c}{\tabularCenterstack{c}{Any outdated \\ dependencies}} & \multicolumn{1}{>{\columncolor{green!40}}c}{\tabularCenterstack{c}{Whether the package has any outdated dependencies.}} & \multicolumn{1}{>{\columncolor{green!40}}c}{\tabularCenterstack{c}{Ensure that all dependencies of the \\ malicious package are up-to-date.}} \\
    \midrule
    \multicolumn{1}{>{\columncolor{green!40}}c}{\tabularCenterstack{c}{Is deprecated}} & \multicolumn{1}{>{\columncolor{green!40}}c}{\tabularCenterstack{c}{Whether the package is deprecated. \\ If so, this metric is set to -5, else to 0}} & \multicolumn{1}{>{\columncolor{green!40}}c}{\tabularCenterstack{c}{Not relevant: it is generally under control of the attacker.}} \\
    \midrule
    \multicolumn{1}{>{\columncolor{green!40}}c}{\tabularCenterstack{c}{Is unmaintained}} & \multicolumn{1}{>{\columncolor{green!40}}c}{\tabularCenterstack{c}{Whether the package is unmaintained. \\ If so, this metric is set to -5, else to 0}} & \multicolumn{1}{>{\columncolor{green!40}}c}{\tabularCenterstack{c}{Not relevant: it is generally under control of the attacker.}} \\
    \midrule
    \multicolumn{1}{>{\columncolor{green!40}}c}{\tabularCenterstack{c}{Is removed}} & \multicolumn{1}{>{\columncolor{green!40}}c}{\tabularCenterstack{c}{Whether the package has been removed from the ecosystem. \\ If so, this metric is set to -5, else to 0}} & \multicolumn{1}{>{\columncolor{green!40}}c}{\tabularCenterstack{c}{Avoid detection by the maintainers of the ecosystem, \\ e.g., by using obfuscation techniques to hide the malicious code.}} \\
    \bottomrule
    \end{tabular}
    \end{adjustbox}
    \vspace{+0.1cm}
    \label{table:metrics}
\end{table*}

\subsection{SourceRank Overview}
SourceRank is a state-of-the-art ranking score developed by Libraries.io to evaluate the popularity and trustworthiness of open-source packages~\cite{Sun2023SoftwareProvenance,Saini2020SoftwareMetrics}.
It consists of an integer score that is computed by summing up the values of 18 metrics, which are described in~\autoref{table:metrics}.

As there isn't any documentation page on SourceRank, the description of the metrics is the result of our hands-on experience using it. The Libraries.io website only offers a brief overview of the metrics (and their value) in the SourceRank breakdown webpage for a given package. As an example, \autoref{fig:pandas_sr} (left) shows the SourceRank breakdown page for the \texttt{pandas} package\footnote{\url{https://libraries.io/pypi/pandas/sourcerank}}.
A noteworthy observation is that 5 out of the 18 metrics, namely \textit{All pre-releases}, \textit{Any outdated dependencies}, \textit{Is deprecated}, \textit{Is unmaintained}, and \textit{Is removed} (highlighted in red in~\autoref{table:metrics}),
are not shown in the SourceRank breakdown webpage, but are provided by the Libraries.io API~\cite{librariesio_api} and used to compute the SourceRank score.
As visible in \autoref{fig:pandas_sr}, the SourceRank breakdown page (on the left) of \texttt{pandas} does not show the values for the 5 metrics, which are instead provided in the JSON report (on the right) obtained through the Libraries.io API.

Such inconsistency, together with the lack of documentation, can lead to a misinterpretation of SourceRank, as users may not be aware of the existence of these metrics or their impact on the overall score.

\begin{figure}[!t]
    \centering
    \includegraphics[width=\columnwidth]{figures/sourcerank_pandas_new.pdf}
    \caption{SourceRank score for the \texttt{pandas} package provided 
    on the Libraries.io webpage (left)
    and obtained through the Libraries.io API as JSON (right).
    }
    \label{fig:pandas_sr}
\end{figure}

\subsection{Threat Model}
\label{sec:threat_modeling}
We present hereafter a threat model of the SourceRank score according to the following four criteria inspired by the adversarial machine learning literature~\cite{biggio2018wild}.
\myparagraphlb{Goal.} The goal of an attacker is to create a malicious package that maximizes its SourceRank score, hence masquerading it as a trustworthy one.
\myparagraphlb{Knowledge.} We assume a \emph{white-box} scenario: the attacker has full knowledge of SourceRank, including its metrics and how they are computed.
\myparagraphlb{Capability.} The attacker has the capability to create a new package on PyPI (or, in general, any other ecosystem that uses the SourceRank score).
\myparagraphlb{Strategy.}
The attacker aims to manipulate the SourceRank metrics to increase the SourceRank score of the malicious package.

To this end, for each metric, we identified several evasion approaches that can be exploited by the attacker to manipulate the corresponding metric (see \autoref{table:metrics}).

\mysubparagraph{Basic info present.} The attacker can create a malicious package with:
\begin{itemize}
    \item a brief description. To make it more appealing to users, the attacker can use a description that is similar to well-known packages (e.g., by using popular buzzwords or appealing sentences);.
    \item a URL that either points to a legitimate package (named in this paper as \emph{URL confusion} attack), or to a new repository created by the attacker (even not containing any malicious code).
    This attack would be also effective if the URL points to a (well-known) website, potentially related to the topic of the package to make it more appealing to users;
    \item keywords that are commonly used by users to search for packages (not necessarily related to the package itself), such as those related to popular legitimate packages or trending topics in the community.
\end{itemize}

All the above techniques can be leveraged to increase the SourceRank of the malicious package to 1 
to make it look like a well-known and trustworthy package.
The key point is that there is no verification of the provided information, hence the attacker can provide any information that is appealing to users.

\mysubparagraph{Source repository present.} The attacker can provide a URL that points to a legitimate source code repository (e.g., a GitHub repository) to obtain a value of 1.
This can be achieved either by referencing an existing one (e.g., from a well-known package) that is also potentially unrelated to the malicious package, or
by creating a new repository on a popular platform (e.g., GitHub, GitLab).
We refer to the first approach as \emph{URL confusion}~\cite{saric2024} as it consists in exploiting the URL of a legitimate package.
It is worth noting that, as already explained for the \emph{Basic info present} metric, the Libraries.io platform does not verify the legitimacy of the provided repository URL.
Hence, the attacker can provide any URL that points to a legitimate source code repository.
Finally, as for the second scenario, to increase the legitimacy appeal when using the URL of a newly created repository, the attacker can also use \emph{typosquatting} for the package name or additional evasion techniques such as
alternate spelling or familiar term abuse~\cite{Ladisa2023SoK,Neupane2023Usenix}.

\mysubparagraph{Readme present.} The attacker can simply provide a README file in the related repository to set this metric to 1.
If the attacker leverages the \emph{URL confusion} attack, the README file can be the one in the related repository of the legitimate package.
On the other hand, if the attacker creates a new repository, the attacker can create a new one that is similar to the README file of a well-known package to make the README file more appealing.

\mysubparagraph{Has multiple versions.} The attacker can create multiple releases of the malicious package to set this metric to 1.
For instance the attacker can publish the first release of the package (or few initial releases) with legitimate code and then update it with malicious code in subsequent releases.
This can be particularly effective if the attacker initially provides a legitimate package to gain the user trust, and then injects malicious code in subsequent releases.

\mysubparagraph{Follows SemVer.} The attacker can follow the Semantic Versioning (SemVer)~\cite{semver} specification to set this metric to 1.

\mysubparagraph{Recent release.} The attacker can simply create a new release of the malicious package every six months to set this metric to 1.
The main challenge for the attacker is to create a new release that is not detected as malicious by the maintainers of the ecosystem.
To this end, the attacker can leverage several obfuscation techniques to hide the malicious code such as encryption and obfuscation of strings~\cite{Ladisa2023SCORED,Collberg2002Obfuscation}.

\mysubparagraph{Not brand new.} This metric is set to 1 if the package has existed for at least six months.
From an attacker's perspective, this means that the malicious package must remain undetected for at least six months after its creation.
To achieve this, the attacker can create an initial version of the package that is legitimate and does not contain any malicious code, and then update it with malicious code after six months.
The main drawback is that the attacker has to wait for six months before being able to inject the malicious code.
On the other hand, as discussed above for the \emph{Recent release} metric, the attacker can leverage several obfuscation techniques to hide the malicious code.

\mysubparagraph{1.0.0 or greater.} The attacker can create a release with a version greater or equal to 1.0.0 to set this metric to 1.

\mysubparagraph{Dependent packages.} This metric is related to the number of projects (\ie, packages available in the ecosystem) that depend on the malicious package.
To obtain a high value for this metric, the attacker can create many additional packages (not necessarily malicious) that depend on the malicious one.
It is worth noting that this solution may not be straightforward as it requires effort from the attacker to create new packages.

\mysubparagraph{Dependent repositories.} This metric depends on the number of repositories that are provided for packages depending on the malicious one.
Thus, is strictly related to the \emph{Dependent packages} metric.
To obtain a high value for this metric, the attacker can extend the previous approach by also creating a repository for each dependent package created.

\mysubparagraph{Stars.} This metric is related to the number of stars received by the repository of the package.
To boost this metric, the attacker can hijack a legitimate repository (\ie, providing in the malicious package a URL that points to a legitimate repository as discussed for the \emph{Source repository present} metric), thus inheriting the stars of the legitimate package exploiting its popularity.
Alternatively, the attacker can create a new repository for the malicious package and try to gain stars by promoting it on social media or by creating fake accounts that star the repository.

\mysubparagraph{Contributors.} This metric is related to the number of contributors to the package and is obtained from the repository provided in the package.
Similarly to the \emph{Stars} metric, the attacker can leverage the repository of a legitimate package, thus inheriting its number of contributors, or create a new repository and fake accounts added as contributors.

\mysubparagraph{Subscribers.} This metric is related to the number of subscribers to the package on Libraries.io.
To obtain a high value for this metric, the attacker can create fake accounts on Libraries.io, which subscribe to the package.

\mysubparagraph{All pre-releases.} This metric is set to -1 (\ie, it penalizes the SourceRank) if the package has all pre-release versions, such as alpha or beta releases, otherwise it is set to 0.
Hence, the attacker can simply avoid creating pre-release versions for the malicious package.

\mysubparagraph{Any outdated dependencies.} This metric is set to -1 (\ie, it penalizes the SourceRank) if the package has any outdated dependencies, 0 otherwise.
To avoid having this metric set to -1, the attacker can simply ensure that all dependencies of the malicious package are up-to-date, or if possible, avoid dependencies at all.

\mysubparagraph{Is deprecated.} This metric is set to -5 (\ie, it heavily penalizes the SourceRank) if the package is deprecated.
It is worth noting that even though this metric is not available on the Libraries.io website, we found out that it is set to -5 if the package is reported as deprecated in the ecosystem.
As for PyPI, since the status is set directly by the package maintainer, this metric is not useful for trustworthiness evaluation, as it is under the control of the attacker.

\mysubparagraph{Is unmaintained.} This metric is set to -5 (\ie, it heavily penalizes the SourceRank) if the package is unmaintained.
Similarly to the \emph{Is deprecated} metric, this metric is not documented on the Libraries.io website, hence we assume it is set to -5 if the package is reported as unmaintained in the ecosystem.
Moreover, also this metric is under the control of the attacker, as it is set by the package maintainer.

\mysubparagraph{Is removed.} This metric is set to -5 (\ie, it heavily penalizes the SourceRank) if the package has been removed from the ecosystem.
To avoid this, the attacker must ensure that the malicious package is not removed from the ecosystem, \ie, that is not detected as malicious by the maintainers of the ecosystem.
To this end the attacker can leverage obfuscation techniques to hide the malicious code and ensure that the malicious package respects the policies of the ecosystem.

\myparagraphlb{Summary.}
We provide a summary of the manipulation strategies that can be exploited by the attacker to increase the SourceRank of a malicious package in~\autoref{table:metrics}.
It is important to note that SourceRank is based on a set of metrics that can be easily manipulated by the attacker.
Hence, only relying on the SourceRank score to evaluate the trustworthiness of a package is not sufficient and can lead to a false sense of security.
Moreover, by just leveraging a URL that points to a legitimate repository, the attacker can influence several metrics at once: \emph{Basic info present}, \emph{Source repository present}, \emph{Readme present}, \emph{Stars}, and \emph{Contributors}.
In particular the attacker can exploit the popularity of the legitimate repository to obtain a high SourceRank score, as well-known projects often have a high number of stars and contributors.
We will refer to this approach as \emph{URL confusion} in the rest of the paper and will show that it is exploited in the wild by several malicious packages.
On top of it, the attacker can easily manipulate other metrics such as \emph{Has multiple versions}, \emph{Recent release}, \emph{1.0.0 or greater}, \emph{All pre-releases}, \emph{Any outdated dependencies} to further increase the SourceRank score of the malicious package.
As for the remaining metrics, we argue that \emph{Dependent packages}, \emph{Dependent repositories}, \emph{Subscribers} and \emph{Not brand new}
are the most difficult to manipulate, as they require more effort from the attacker (e.g., creating many additional packages that import the malicious package as dependency, or creating fake accounts on Libraries.io to subscribe to the package).
Finally, we note that the \emph{Is deprecated} and \emph{Is unmaintained} metrics are not useful for trustworthiness evaluation, as they are under the control of the attacker.
\section{Experiments}
\label{sec:experiments}
In this section, we present the experiments conducted to analyze how the SourceRank score behaves for benign and malicious packages in the PyPI ecosystem, and to assess the prevalence of the evasion technique based on \emph{URL confusion}.
We first introduce the research questions we aim to answer in this work, then we describe the datasets we used for our analysis, and finally we present the results of our experiments.

\subsection{Research Questions}

\myparagraph{[RQ.1] SourceRank distributions analysis} -- 
How do the SourceRank distributions of benign and malicious packages compare within the PyPI ecosystem?

\myparagraph{[RQ.2] SourceRank effectiveness to discriminate benign from malicious packages.} -- 
How effective is the SourceRank to discriminate benign from malicious packages in the PyPI ecosystem?

\myparagraph{[RQ.3] Presence of \emph{URL confusion} in the wild} -- How many malicious packages leverage the \emph{URL confusion} evasion approach?
\subsection{Datasets}
For our experiments we adopted two different datasets, namely MalwareBench~\cite{malwarebench} and a real-world dataset of Python packages collected from PyPI, 
which are described in the following.

\myparagraphlb{MalwareBench.}
\emph{MalwareBench} is a state-of-the-art dataset collected by Zahan \etal~\cite{malwarebench}, which includes 3,190 unique malicious and 3,368 unique benign Python packages.
We use this dataset as baseline to analyze the distribution of the SourceRank score in the PyPI ecosystem, 
as well as to understand if the \emph{URL confusion} evasion approach has been exploited by malicious packages in the past.
We collected the SourceRank scores of the packages in the MalwareBench dataset on May 9th, 2025 using the Libraries.io API~\cite{librariesio_api}.
We filtered out 39 packages (20 malicious and 19 benign) whose SourceRank was not available (e.g., when a package is removed from the ecosystem by its maintainer), and we took the latest release of each package, 
resulting in a total of 3,170 malicious and 3,349 benign packages.

\myparagraphlb{Real-world dataset.}
We built a real-world dataset by collecting all packages uploaded to PyPI over 80 days (from October 2nd, 2024 to December 21st, 2024),
using the PyPI feeds for newly uploaded packages\footnote{\url{https://pypi.org/rss/packages.xml}} and releases\footnote{\url{https://pypi.org/rss/updates.xml}}.
The dataset contains 48,711 unique packages (214 malicious and 48,503 benign) and 122,398 releases. 
We collected the ground truth labels of the packages by leveraging the Open Source Security Foundation (OpenSSF) repository of malicious packages~\cite{openssf_db_malware}
and the private PyPI database~\cite{pypi_db_malware}, a private GitHub repository\footnote{access is granted by the PyPI security team} including the packages reported as malicious by the PyPI maintainers and the community.
We collected the labels five days after the end of the collection period and rechecked them five months later, confirming that they were still valid.
As for the MalwareBench dataset, we collected the SourceRank scores for the packages on the May 9th, 2025, using the Libraries.io API~\cite{librariesio_api}.
Finally, we filtered out 51 packages (4 malicious and 47 benign) whose SourceRank was not available, and we took the latest release of each package,
resulting in a total of 210 malicious and 48,456 benign packages.

\subsection{Experimental Setup}
All experiments were conducted on an Ubuntu 22.04.6 LTS server equipped with an Intel Xeon Platinum 8160 CPU @ 2.10 GHz (64 cores) and 256 GB of RAM.
To collect the SourceRank scores of the packages, we used the API provided by Libraries.io~\cite{librariesio_api}.
For the analysis of the datasets, we used Python 3.11.9 and the \texttt{pandas} library~\cite{pandas}.

\begin{figure*}[!t]
\centering
\begin{subfigure}{\columnwidth}
    \centering
    \includegraphics[width=\textwidth]{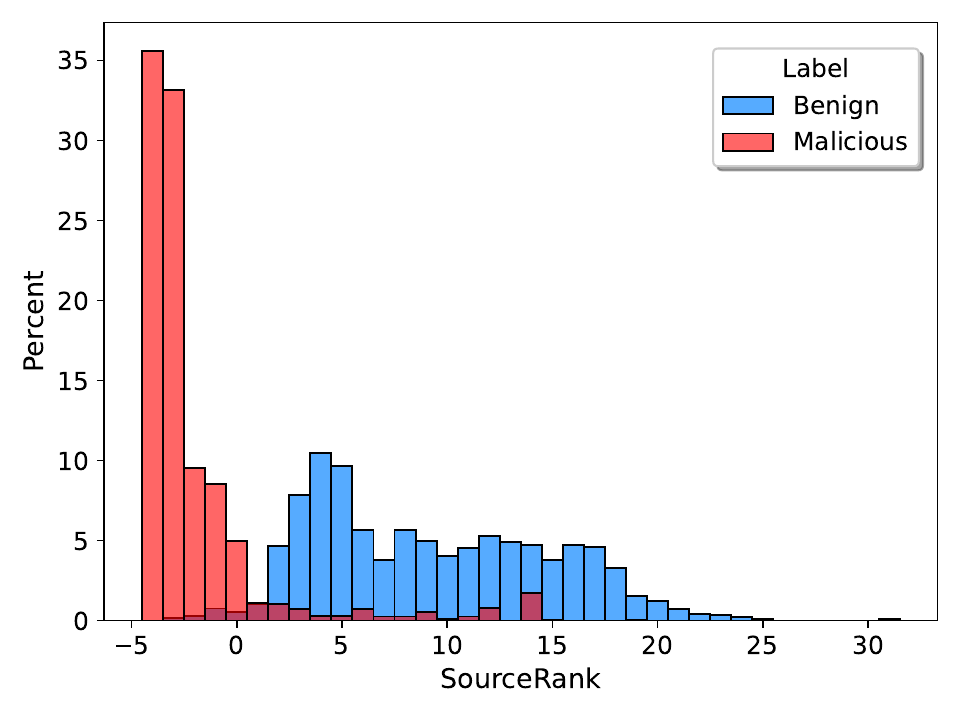}
    \caption{MalwareBench dataset}
    \label{fig:sr_malwarebench}
\end{subfigure}
\begin{subfigure}{\columnwidth}
    \centering
    \includegraphics[width=\textwidth]{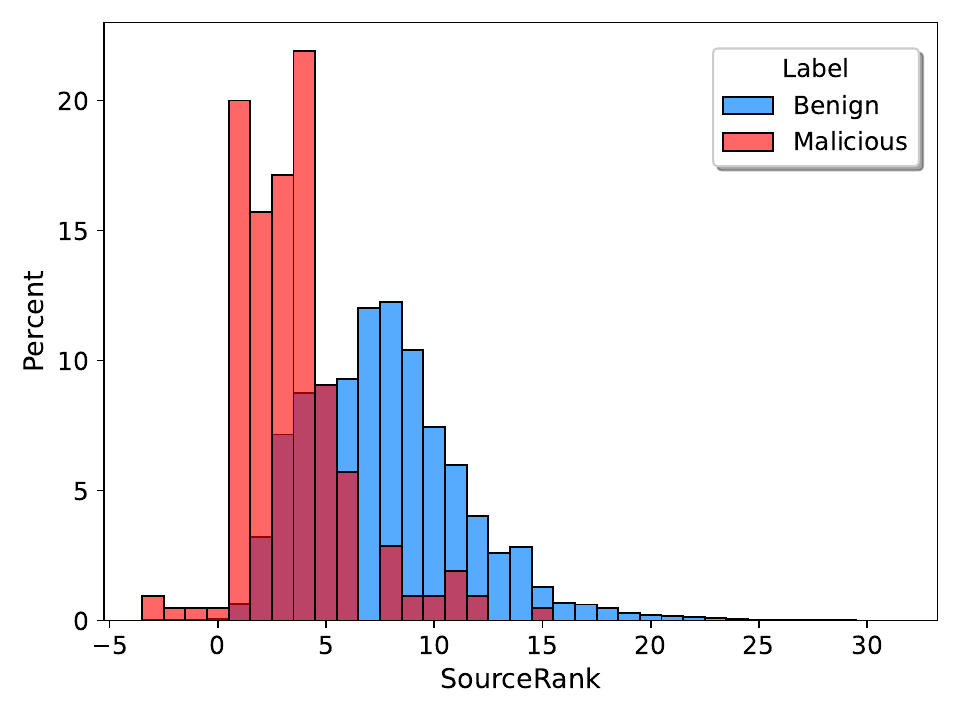}
    \caption{Real-world dataset}
    \label{fig:sr_real_world}
\end{subfigure}
\caption{SourceRank distributions of benign (blue) and malicious (red) packages for (a) MalwareBench and (b) real-world datasets.
The other color in the plot (purple), is given by the overlapping between the blue and red distributions.
}
\label{fig:source_rank_distribution}
\end{figure*}

\begin{figure}[!t]
    \centering
    \includegraphics[width=\columnwidth]{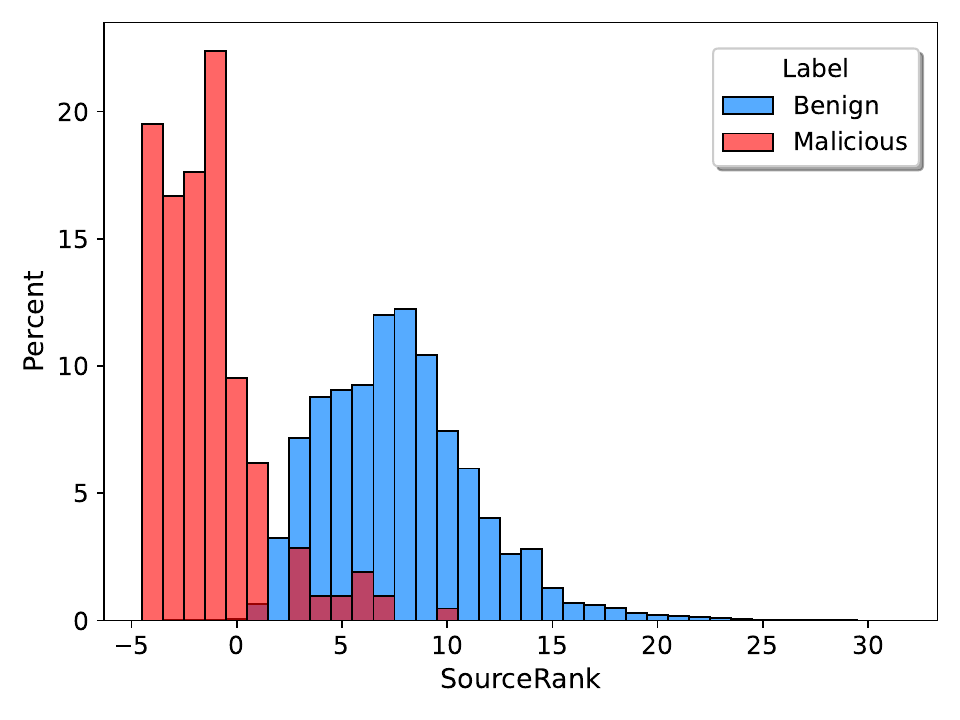}
    \caption{SourceRank distributions of benign and malicious packages for the real-world dataset after updating the \emph{Is Removed} metric.
    }
    \label{fig:sourcerank_live_updated}
\end{figure}

\begin{figure}[!t]
    \centering
    \includegraphics[width=\columnwidth]{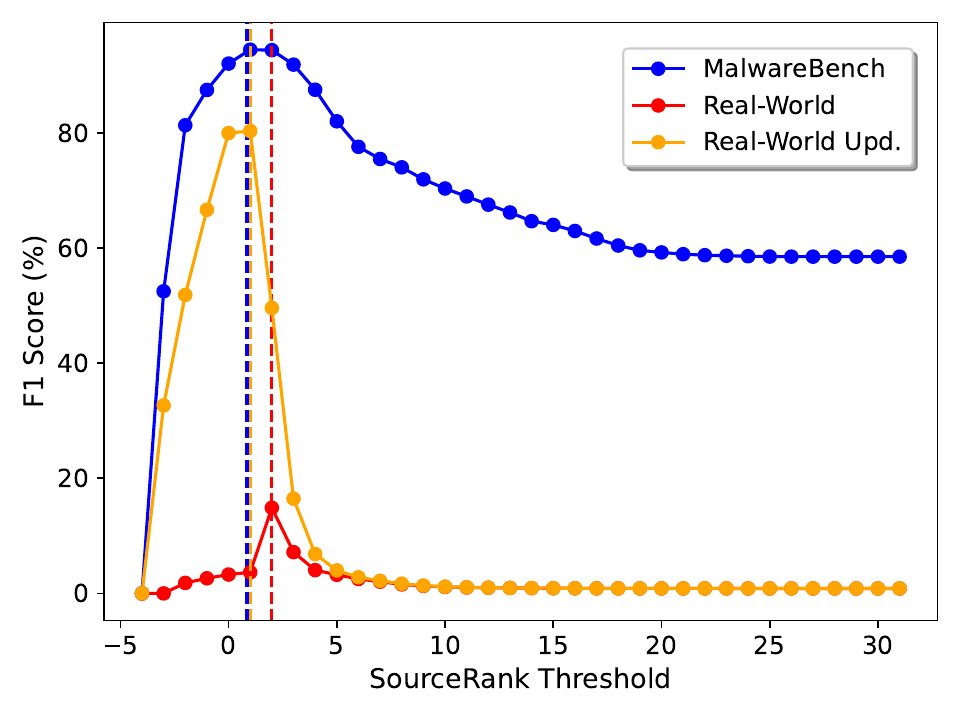}
    \caption{Tuning of the SourceRank threshold for detection of malicious packages on the target datasets: MalwareBench (blue) 
    and real-world (red for the original SourceRank, yellow for the one obtained after updating the \emph{Is Removed} metric).
    }
    \label{fig:sourcerank_detection}
\end{figure}

\begin{table}[t]
    \centering
    \caption{Statistics of the SourceRank score for MalwareBench and real-world datasets.}
    \begin{adjustbox}{max width=\columnwidth}
    \begin{tabular}{c c c c c}
        \toprule
        \textbf{Metric} & \multicolumn{2}{c}{\textbf{MalwareBench}} & \multicolumn{2}{c}{\textbf{Real-world}} \\
        \cmidrule(lr){2-3}  \cmidrule(lr){4-5}
        & \textbf{Benign} & \textbf{Malicious} & \textbf{Benign} & \textbf{Malicious} \\
        \toprule
        Min         &  -3    &  -4     &  -3   &  -3    \\
        Max         &  31    &  19     &  31   &  15    \\
        Mean $\pm$ Std   &  9.21 $\pm$ 5.54  &  -2.09 $\pm$ 3.35 &  7.75 $\pm$ 3.64  &  3.46 $\pm$ 2.57 \\
        Median      &  8     &  -3     &  7    &  3     \\
        \bottomrule
    \end{tabular}
    \end{adjustbox}
    \vspace{+0.1cm}
    \label{tab:sourcerank_stats}
\end{table}

\subsection{Results}

\myparagraphlb{SourceRank distribution analysis.} \autoref{fig:source_rank_distribution} shows the distribution of the SourceRank score, expressed as the percentage of packages for each score value from -5 to 32, for benign and malicious packages in both datasets.
We report the statistics (min, max, mean and std. deviation) in Table~\ref{tab:sourcerank_stats}.

While for the MalwareBench dataset (cf. \autoref{fig:sr_malwarebench}) the distributions for benign and malicious packages are well-separated with little overlap (the mean SourceRank of benign packages is 9.21, while for malicious ones is -2.09), in a real-world dataset, the distributions are more similar and tend to overlap (mean SourceRank of 7.75 for benign packages and 3.46 for malicious ones).
It is worth noting that while the mean SourceRank of benign packages remains similar in both datasets (9.21 vs. 7.75), for malicious packages the mean SourceRank significantly increases from -2.09 in the MalwareBench dataset to 3.46 in the real-world dataset.

We further investigated this difference and found that 91.8\% of the malicious packages in the MalwareBench dataset that have a SourceRank lower or equal to 0 and 95.5\% have the \emph{Is Removed} metric set to -5, which indicates that the package has been removed from PyPI.
On the other hand, in the real-world dataset, only 2.4\% of the malicious packages have a SourceRank lower or equal to 0 and only 2.86\% have the \emph{Is Removed} metric set to -5.
The last finding is quite surprising: although all malicious packages in the real-world dataset have been removed from PyPI (which we confirmed by querying the platform), the SourceRank scores do not always reflect this and have a value of 0 instead of -5 for the \emph{Is Removed} metric for 204 out of the 210 malicious packages of the real-world dataset.
In fact, only 6 (2.9\%) packages are reported as removed by Libraries.io despite we collected the SourceRank scores five months after the end of the collection period, and most of the packages, 
\ie, 177 (84.3\%) out 210 malware packages\footnote{the remaining 33 packages are included only in the PyPI private database~\cite{pypi_db_malware}}, were already reported as malicious by the public OpenSSF database~\cite{openssf_db_malware}.
Hence, the information about the maliciousness of these packages was already available online but is not reflected by the SourceRank.

We also investigated how the distributions change if we set the \emph{Is Removed} metric to the correct value of -5 for all the malicious packages in the real-world dataset (see \autoref{fig:sourcerank_live_updated}).
We found that the SourceRank distributions of the malicious packages are significantly shifted to the left, with a mean SourceRank of -1.40 compared to 3.46 before, much closer to the MalwareBench distribution.
This confirms that a major cause of the misalignment between the distributions for the MalwareBench and real-world datasets is that SourceRank does not promptly reflect the fact that malicious packages have been removed from PyPI.

\vspace{0.2cm}
\takeawaysNoTitle{
\textbf{[RQ.1]} The SourceRank distributions of benign and malicious packages differ only on historical data (MalwareBench), due to the metric \emph{Is Removed} that is set to -5 for all the malicious packages.
In the real-world dataset, the distributions overlap significantly as SourceRank does not promptly reflect the real status of malicious packages that have been removed from PyPI.
}

\myparagraphlb{SourceRank effectiveness to discriminate benign from malicious packages.}
We analyze here the effectiveness of SourceRank in discriminating benign from malicious packages.
To this end, we aim to find the best threshold for SourceRank that maximizes the F1-score (defined as the harmonic mean of precision and recall) on the MalwareBench dataset and validate it on the real-world dataset.
The results, shown in \autoref{fig:sourcerank_detection}, confirm the findings of RQ1: while for the MalwareBench dataset (blue), a threshold of 1 achieves the best F1-score of 94.5\%,
with the same threshold on the real-world dataset (red), the F1-score drops to 3.66\%.
It is worth noting that even if we tune the threshold on the real-world dataset, the best F1-score is only 14.87\% (with a threshold of 2), which is still significantly lower than the performance on the MalwareBench dataset and impractical for real-world use.
On the other hand, if we update the \emph{Is Removed} metric to -5 for all the malicious packages in the real-world dataset (green) and apply the same threshold tuned on the MalwareBench dataset, 
the F1-score increases to 80\%.
These results confirm again that the SourceRank score is not effective in discriminating malicious packages in a real-world scenario, mainly because it does not reflect whether a package has been removed. 
It is worth remarking that being able to discriminate malicious packages once they are removed is of little interest, as the package is no longer available on PyPI and cannot be downloaded by users.
This result highlights the unreliability of the SourceRank score in selecting benign packages among those available on PyPI.

\vspace{0.2cm}
\takeawaysNoTitle{
\textbf{[RQ.2]}
SourceRank is not effective in discriminating benign from malicious packages in a real-world scenario 
because it does not reflect whether a package has been removed from PyPI.
Moreover, it is not reliable in selecting benign packages among those available on PyPI.
}

\begin{table}[!t]
    \centering
    \caption{Analysis of the \emph{URL confusion} evasion approach for the real-world dataset.}
    \begin{adjustbox}{max width=\columnwidth}
    \begin{tabular}{c c}
        \toprule
        \textbf{Repository} & \textbf{Counting} \\
        \toprule
        \href{https://github.com/pypa/sampleproject}{\texttt{pypa/sampleproject}} & 3 \\
        \href{https://github.com/CorwinDev/Discord-Bot}{\texttt{CowinDev/Discord-Bot}} & 1 \\
        \href{https://github.com/encode/httpx}{\texttt{encode/httpx}} & 1 \\
        \href{https://github.com/fake-useragent/fake-useragent}{\texttt{fake-useragent/fake-useragent}} & 1 \\
        \href{https://github.com/psf/requests}{\texttt{psf/requests}} & 1 \\
        \href{https://github.com/cuongitl/python-bitget}{\texttt{cuongitl/python-bitget}} & 4 \\
        others & 4 \\
        \midrule
        Total & 15 (7\%) \\
        \bottomrule
    \end{tabular}
    \end{adjustbox}
    \vspace{+0.1cm}
    \label{tab:url_jacking_analysis}
\end{table}

\myparagraphlb{Presence of \emph{URL confusion} in the wild.}
In this RQ, we analyze the presence of \emph{URL confusion} on both datasets.
To this end, we first check if the packages in the datasets have the \emph{Source Repository Present} metric set to 1, which means that the package has a source repository available.
Then, we manually check if the URL of the package points to a repository associated with a legitimate package\footnote{We consider a repository legitimate if it is associated with a package that is not malicious according to the OpenSSF database~\cite{openssf_db_malware} and the PyPI private database~\cite{pypi_db_malware}.}.
We found that this evasion approach is present in 134 (4.2\%) malicious packages of the MalwareBench dataset, while it is exploited in 15 (7.0\%) malicious packages of the real-world dataset.
This shows that, even though this evasion approach is not widely used, it is present in the wild with an increasing trend (from $\sim$4\% in the MalwareBench dataset to 7\% in the real-world dataset).

Among the most common legitimate repositories exploited by the attackers, we found two different malware campaigns (3 packages in total) that exploit the \texttt{sampleproject} repository\footnote{\url{https://github.com/pypa/sampleproject}},
which is a template repository provided by the PyPI maintainers to help developers create new packages.
This template repository has more than 5,000 stars on GitHub and more than 50 contributors, which makes it an attractive target for attackers to exploit.
We also found another campaign made of 2 packages, namely \texttt{discordbotpresence-0.6.7}\footnote{\url{https://libraries.io/pypi/discordbotpresence}} and \texttt{discordbotstatus-0.6.7}\footnote{\url{https://libraries.io/pypi/discordbotstatus}},
that exploit the \texttt{Discord-Bot}\footnote{\url{https://github.com/CorwinDev/Discord-Bot}} and \texttt{httpx}\footnote{\url{https://github.com/encode/httpx}} repositories, respectively,
the latter of which is a popular HTTP client for Python with more than 14,000 stars on GitHub and more than 200 contributors.
In addition, we found a package named \texttt{fake-usreagent}\footnote{\url{https://libraries.io/pypi/fake-usreagent}} that leverages typosquatting to impersonate the \texttt{fake-useragent}\footnote{Note that the attacker swapped the \textbf{e} with \textbf{r} in the \textbf{useragent} word: fake-us\textbf{er}agent $\Longrightarrow$ fake-us\textbf{re}agent} package and
leveraged the URL pointing to its repository\footnote{\url{https://github.com/fake-useragent/fake-useragent}}.

We also found additional campaigns that exploit popular repositories through similar typosquatting techniques and \emph{URL confusion}.
For instance, the \texttt{frexco-pip-requests}\footnote{\url{https://libraries.io/pypi/frexco-pip-requests}} package exploits the well-known \texttt{requests} package,
which is a popular HTTP library for Python with more than 50,000 stars on GitHub and more than 600 contributors.
It is worth noting that its name has been chosen to resemble the name of the \texttt{requests} package by using the prefix augmentation technique~\cite{Neupane2023Usenix},
a common approach for dependency confusion attacks~\cite{Neupane2023Usenix}.
Finally, we also found a campaign made of 4 packages obtained by suffix augmentation~\cite{Neupane2023Usenix}, namely 
\codeinline{python-bitget-api-3.3.5}\footnote{\url{https://libraries.io/pypi/python-bitget-api}},
\codeinline{python-bitget-connect-0.3.9}\footnote{\url{https://libraries.io/pypi/python-bitget-connect}},
\codeinline{python-bitget-request-4.9.5}\footnote{\url{https://libraries.io/pypi/python-bitget-request}}, and 
\codeinline{python-bitget-wrapper- 0.3.7}\footnote{\url{https://libraries.io/pypi/python-bitget-wrapper}},
which exploit the \codeinline{bitget-api}\footnote{\url{https://github.com/cuongitl/python-bitget}} repository, a Python library for cryptocurrency trading on the Bitget exchange.

Finally, we analyzed the SourceRank of the packages that leverage this evasion technique and found that the average SourceRank score is 10, while the maximum is 15,
which is particularly high compared to the average SourceRank of malicious (3.46) and, above all, benign (7.75) packages in the real-world dataset.
Moreover, all of them also leverage other evasion techniques described in the threat modeling (see Section~\ref{sec:sr_threat_modeling})
to manipulate other metrics that are very easy to manipulate, such as \emph{Follows SemVer} or \emph{Readme present}.
This confirms that \emph{URL confusion}, combined with other evasion approaches, can effectively increase the SourceRank of a malicious package,
hence masquerading it as a trustworthy package.

\vspace{0.2cm}
\takeawaysNoTitle{
\textbf{[RQ.3]}
The \emph{URL confusion} evasion approach is an emerging technique that is exploited by 7\% of the malicious packages in the real-world dataset (compared to 4.2\% in the MalwareBench dataset).
Combined with other evasion techniques, this approach can result in very high SourceRank scores (up to 15),
hence allowing attackers to effectively masquerade a malicious package as a trustworthy one.
}
\section{Conclusions and Future Work}
\label{sec:conclusions}

\myparagraphlb{Conclusions.}
The software supply chain requires nowadays robust and reliable scoring systems to assess the trustworthiness of open-source projects and packages.
In this work, we focused on SourceRank, a well-known score developed by Libraries.io for ranking and evaluating the popularity of open source packages.
Despite being adopted in several recent studies~\cite{latendresse2025robustllmgeneratedlibraryimports,latendresse2025softwarelibrarian,latendresse2024chatgptgoodsoftwarelibrarian,Jia2024EmpiricalStudy,Sun2023SoftwareProvenance,Ladisa2023ACSAC,Ye2022ICSE},
its reliability against evasion attacks aiming at increasing the score of a malicious package has never been thoroughly analyzed.
To address this open issue, we first performed a threat modeling of the SourceRank score and then analyzed its effectiveness in discriminating benign from malicious packages in the PyPI ecosystem.

We identified several evasion techniques against the metrics that contribute to the SourceRank score, and, among these, we discovered
a new evasion technique named \emph{URL confusion}, which consists in leveraging a URL that points to a legitimate package and can be exploited to manipulate 5 out of the 18 metrics that contribute to the score.

We performed a study on the PyPI ecosystem by analyzing the SourceRank distribution of benign and malicious packages from two different datasets: the state-of-the-art MalwareBench dataset, and a large dataset of 122,398 packages collected daily from PyPI over 80 days.
Our analysis revealed that SourceRank is not reliable in a real-world scenario, as it does not promptly reflect the removal of packages from PyPI. 
Malicious packages in our real-world dataset that had been removed from PyPI were still not recognized as removed five months after their collection, despite
their maliciousness being publicly available on several advisories such as the OpenSSF malware database~\cite{openssf_db_malware}.

Finally, we analyzed the presence of \emph{URL confusion} in our datasets, and showed that 
it is an emerging attack technique (from 4.2\% of the packages in MalwareBench to 7.0\% of the packages in the real-world dataset) and that
malicious packages can achieve a significantly high SourceRank score (up to 15), especially when combined with other techniques.

By empirically showing the limitations of SourceRank, our study raises awareness about the false sense of security it may create and helps to improve the existing scoring systems for evaluating the security posture of open source projects.

\myparagraphlb{Future Work.}
The analysis of SourceRank can be further extended to other ecosystems such as npm.
Moreover, the same threat modeling can be applied to other scoring systems such as OpenSSF Scorecard~\cite{scorecard_paper,OpenSSFScorecard}.
In addition, to counter the identified limitation of SourceRank about package removal status,
some possible direction are periodically checking the removal status though the repository APIs,
or leveraging reliable community reports such as the OpenSSF malware database~\cite{openssf_db_malware}.
Finally, to mitigate the \emph{URL confusion} evasion technique, a possible direction is to implement a robust verification mechanism that checks whether the URL provided in a package's metadata actually points to the package's repository or project page.
This can be achieved by combining several verification methods, such as checking for the repository structure (same files and directories), compare release artifcats available in the repository with those in the package registry (if available) and
time-based correlation of package releases and repository releses.

\printbibliography

\end{document}